# A Privacy-Preserving Data Inference Framework for Internet of Health Things Networks

James Jin Kang, Mahdi Dibaei, Gang Luo, Wencheng Yang and Xi Zheng

***Abstract*—Privacy protection in electronic healthcare applications is an important consideration due to the sensitive nature of personal health data. Internet of Health Things (IoHT) networks have privacy requirements within a healthcare setting. However, these networks have unique challenges and security requirements (integrity, authentication, privacy and availability) must also be balanced with the need to maintain efficiency in order to conserve battery power, which can be a significant limitation in IoHT devices and networks. Data are usually transferred without undergoing filtering or optimization, and this traffic can overload sensors and cause rapid battery consumption when interacting with IoHT networks. This consequently poses restrictions on the practical implementation of these devices. As a solution to address the issues, this paper proposes a privacy-preserving two-tier data inference framework – this can conserve battery consumption by reducing the data size required to transmit through inferring the sensed data and can also protect the sensitive data from leakage to adversaries. Results from experimental evaluations on privacy show the validity of the proposed scheme as well as significant data savings without compromising the accuracy of the data transmission, which contributes to energy efficiency of IoHT sensor devices.**

## I. Introduction

The introduction of tracking apps in response to the COVID-19 pandemic have highlighted discussions on the vulnerabilities and potential privacy threats that can be associated with these applications. Private information is vulnerable to being compromised using communication protocols with weak security, such as Bluetooth which has been used by some health agencies [1]. Smart home environments emerged with health applications are being prevalent as more homes are being connected to the Internet of Things (IoT) and IoHT networks along with wearable devices. The growing demand for these services will add additional transactions and increase the workload of wireless body area networks (WBAN), which consist of sensors and smartphones. These devices, such as physiological sensors and monitoring devices will be affected by an increased demand in performance and battery power. Sensors in their current capacity do not interact significantly with IoT networks, nor do they have the intelligence to be able to provide data to health networks securely. Instead, these devices are simply passive and only provide data at a regular interval or on-demand due to their hardware size and battery limitations. The use of smartphones to interact with sensors and wearables allows the possibility of these limitations to be overcome by taking advantage of more powerful resources from smartphones. Privacy is a requirement in dealing with health information. Given that this is a new and emerging field, i.e. sensors interacting with IoHT devices, which will use health data, there has been little research to address the privacy concern of sensitive patient data within such a context. It is expected that the volume of traffic and transactions of data requests to sensors in IoT networks will increase as IoT increase in connectivity [2].

In summation, the two critical challenges of a smart home system will be energy efficiency and privacy preservation. To alleviate these problems, we propose a two-tier data inference framework for a smart house care system in which less sensitive information is transferred and encrypted and at the same time energy consumption can be reduced at the wearable devices level (the first tier). Privacy preservation is achieved at the edge servers deployed in each home (the second tier). The first tier infers data processing of sensors to reduce transactions from sensors to smartphones and IoHT networks. Processed encrypted data from wearable devices will be passed to the second tier. To protect the privacy of each resident, the second tier protects data by Laplace noise enabled differential privacy. The two-tier approach is created specifically for IoHT applications where privacy in the underlying sensor data is protected by a privacy-preserving workflow. In these applications, the sensor data is first reduced, and the encrypted sensor data is then transmitted to the edge servers. At the edge servers, differential privacy is used to protect privacy further. This study provides three major contributions: (1) Leveraging model driven prediction, encryption, and data points with edge

J. J. Kang is with Edith Cowan University, Joondalup WA 6027 Australia (e-mail: james.kang@ecu.edu.au).

M. Dibaei is with the Department of Computer Engineering, Islamic Azad University Tabriz Branch, Iran (e-mail: dibayimahdi@yahoo.com).

G. Luo is with Department of computing, Macquarie University, NSW, Australia (e-mail: gangluo96@gmail.com)

W.Yang is with Edith Cowan University, Joondalup WA 6027 Australia (e-mail: w.yang@ecu.edu.au)

Xi Zheng is with the Department of Computing, Macquarie University, NSW, Australia (e-mail: james.zheng@mq.edu.au).

computing to propose a two-tier privacy-preserving IoHT framework. (2) Evaluation of the proposed system in terms of efficiency and privacy preservation with up to 51.3% of savings rate with a good accuracy rate, and (3) Presenting potential application scenarios to benefit from the solution.

## II. RELATED WORK

An IoHT network is defined as an IoT network that includes a personal health device (PHD), which itself is defined in further detail by the IEEE P11073 PHD Work Group. IoHT could include any health devices attached on or within a user's body, and is battery driven with the ability to sense certain health or physiological data. It should also be able to store memory and have capability to communicate wirelessly. To assess the proposed solution with PHD devices and networks, related areas are reviewed including inference systems and privacy-preservation techniques.

### A. Health Inference and Prediction Analytics

Overhead requirements can be reduced by inferring health data on sensor devices, overcoming the need to have a managing device such as a smartphone or other smart device to act as a gateway to connect to the public network.

Engel et al. [3] considered a context aware inference model to process a huge amount of data generated by wireless sensor networks used in logistics operations. Multi-sensor data combined with a context aware inference model delivers relevant information to the user for fast processing and support of large amounts of data, which allows for real-time monitoring of temperature-controlled supply chains. The hybrid context aware model retrieves sensor data and infers this data to produce relevant information. Whilst this may be useful in processing large amounts of data using a context aware inference system, it does not consider inferencing within a situation that requires controlling the volume of data against traffic and information requested from an external party such as IoHT. Zhu et al. [4] put forth a dynamic Bayesian model for averaging in their proposal to develop a high-accuracy prediction analytic method suitable for a large scale IoT application. This method, however, cannot be applied to mHealth data due to the difference in size and type of data between mHealth and IoT networks.

### B. Privacy-Preservation

Privacy-Preservation is critical in many networks such as the cloud, wireless sensor networks and especially in the eHealth environment. There are security and privacy aspects to be considered when transmitting health data to any network. The following is a brief of two types of existing privacy-preserving technologies that are relevant to this study.

*Cryptography based schemes*: Encryption can be defined as an ordered quintet *(P, C, K, E, D),* where *P* is the *plaintexts, C* is the *crypto texts, K* is the *keys, E* is the *encryption functions, D* is the *decryption function.* Pasupuleti et al. [5] proposed a secure privacy-preserving scheme based on probabilistic public key encryption algorithm for securing outsourced data of resource-limited mobile devices. To reduce the computation and communication overhead, the proposed system uses a ranked keyword search which first returns the most relevant files instead of all the files back. Wang et al. proposed a hierarchy attribute-based encryption scheme to secure the shared data using ciphertext-policy attribute-based encryption. An integrated access structure together with some attributes are involved in the encryption. The proposed scheme is proved to be conspicuous efficient with the increasement of the number of files [6]. Wasters [7] proposed an attribute-based encryption. In his solutions, it allows for the data sender to determine the access control policies. A user can decrypt the ciphertext only when the access tree associated with that ciphertext is satisfied by the attribute set which is associated with the private key.

*Differential privacy based schemes*: A major problem with sharing information about a dataset is privacy preservation. Differential privacy is a technique for modifying data in a way that prevents inferring much about private information.

Yin et al. [8] proposed a location privacy-preserving scheme based on the differential privacy strategy for IoT networks. A location information tree model is constructed to express the position dataset. The authors claimed the proposed scheme can achieve higher processing efficiency compared to traditional location privacy protection algorithms. Xu et al. [9] proposed a framework for IoT data analysis called local differential privacy obfuscation, which can ensure that the users' sensitive data will not be exposed when they are aggregated and distilled at the IoT devices. Liu et al. [10] proposed a framework to direct traffic flow from one smart home to anther home gateway prior to sending to the internet, which achieves strong differential privacy and the attacker is unable to link the traffic flow to a specific smart home network.

## III. THE SOLUTION

To reduce the power consumption of IoHT devices and protect sensitive health data generated by IoHT networks, we propose a privacy-preserving two-tier data inference framework. The first tier in this framework involves a data inference algorithm that can reduce the number of redundant or low-value transactions so as to save the power consumption; the second tier protects the sensitive data by using encryption and differential privacy techniques.

### A. The first tier: data reduction using a data inference algorithm

It is unnecessary to consume bandwidth and power resources by sending all available data if there could be a more effective method of reducing the number of original data sent. Therefore, in the first tier, we propose the use of a data inference algorithm which decides only to transmit data if it is significantly different from the previously captured DPs, thus reducing the number of redundant or low-value data transfers [11]. Using this solution, there is a risk that it could not represent the original data and does not properly represent data in certain situations, such as in the case of short interval sampling. To reduce these instances, it is proposed to analyze the differences between the original and inferred data, and apply regular beacons (DPs that are transmitted regardless) into the inferred results such that they act as a framework to reflect the original data as much as possible. Three aspects are considered to assess the results [12] including (1) Efficiency Ratio (ER) of saved (reduced) data volume and actual transmitted data, (2) Savings Ratio (SR) of reduced data and sensed data (%), and (3) Accuracy Ratio (AR) of total value of transmitted data and original data (%).

Variance rate (VR) is used for inferring the selection and subsequent transmission of data. It compares the DP with those directly before and afterwards to screen out DPs which are too similar and do not have to be transmitted, i.e. do not provide new significant information from previous DPs. Different levels of granularity can be applied for VRs, e.g. 1% VR is finer than 10% VR. It can be applied using the formula below.

$$If\ |Vc - Vc1|OR|Vc0 - Vc|) > Vc \times Vr,$$

$$then\ Vx = Vc$$

$$Else\ then\ Vx = Nil,\ where\ Vc = current\ value, Vc0 = previous\ value, Vc1 = next\ value\ and\ Vx = sampling\ value, and\ Vr = variance\ rate$$

When a VR is applied to data, a difference between the graph of inferred data versus the graph of original data will inevitably arise, as depicted in Fig. 1. In this figure, S (Upper) represents the area of this difference or the distorted portion by the inferred values that are less than the original, whilst S (Lower) represents areas of inferred values that are higher than the original. A larger total area of the gap refers to greater data distortion and therefore reducing this gap would be better for accuracy. The formula below depicts the area of upper and lower sides of the inferred graph against the original.

$$Su = \sum_{k=0}^{n} \binom{n}{k} S_n \quad \text{where } S_n = \text{G(S1, S2 ..., Sn)}, \tag{1}$$

similarly

$$S_l = \sum_{k=0}^{n} \binom{n}{k} S_n \quad \text{where } S_n = \text{Y(S1, S2 ..., Sn)}, \tag{2}$$

Total area of the gaps would be presented as below. A larger value means a 'coarser' and higher VR inference has been used relative to a smaller total area which means that a 'finer' and lower VR value has been applied.

The larger the difference ($S_d = |S_u - S_l|$), the farther the result is from the average and less representative of the original trend. However, it is important to note that a smaller difference does not necessarily mean that it represents the original data graph properly – it could however be an indicator of how accurate the inference is to the original along with the gaps instead. For example, a small S value as well as a small $S_d$ suggests it is likely to be closer to the original. These figures in conjunction (i.e. S and $S_d$) can be used to determine how accurate each inference is, whilst the savings or the reduction of DP indicate the efficiency.

A $S$ = 0 would suggest that the inference represents the original data perfectly with no distortion, whilst a $S_d$ = 0 suggests that the inference represents the mean value of the graph despite not representing the original perfectly. Fig. 1 depicts the upper and lower gaps after inference has been applied.

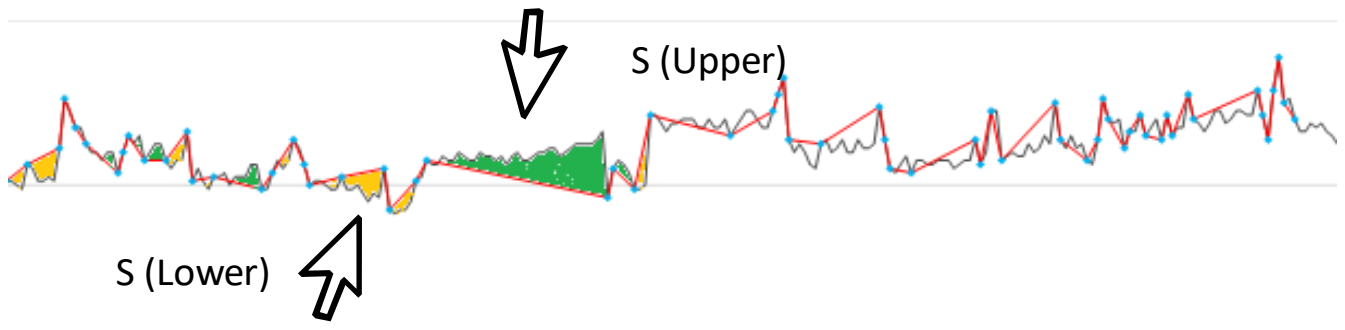

Fig. 1. Differences between the original and inferred value, which shows the areas to be used for accuracy calculation [12]

A consequence of sampling in statistical inference systems is the reduction of DPs, which leads to data size reduction. Increasing VR results in increased savings, however there should be a threshold to ensure accuracy of the result. Privacy preservation is increasingly recognized as a serious concern for IoHT networks where healthcare data is shared, processed and transferred. The degree of privacy preservation in inference systems would be described in what extent that inferred data are different from original data. The sampled data will then be encrypted using symmetric key encryption (SKE), or attribute-based encryption (ABE). As an explanation, ABE is a public key encryption (PKE) technique [13]. The encrypted data are then passed to the second tier.

### *B. The second tier: data protection with differential privacy*

The second tier concerns the protection of sensitive health data created by the IoHT network in the first tier. To protect the privacy of sensitive data in the dataset, removing identifying and personal information such as the user's name, ID, and phone number is insufficient because the remaining data reveals identities in the dataset. Differential privacy is a technique that ensures protection against attackers to infer private information [14]. In the differential privacy algorithms, a randomized function adds a random noise to the true answer to produce a response to a query [15].

**Definition of differential privacy:**

Let D and D' be two neighboring datasets and M a randomized function. M provides ϵ-differential privacy for all sets of O ⊆ Range (M), if it satisfies the following:

$$\frac{\Pr[\, M(D) \in O\,]}{\Pr[\, M(D') \in O]} \leq \exp(\epsilon) \tag{3}$$

It is said that algorithm M provides $\epsilon$-differential privacy protection. It can be seen from the definition of differential privacy that the $\epsilon$ is used to control the probability ratio of the algorithm M to obtain the same output on two adjacent data sets. It reflects the level of privacy protection that M can provide. In practical applications, $\epsilon$ usually takes a small value, such as 0.01, 0.1, or 1n 2, 1n 3. The value of $\epsilon$ should be combined with specific requirements to achieve a balance of safety and the availability of output results. Differential privacy protection can be achieved by adding an appropriate amount of interference noise to the return value of the query function. Adding too much noise will affect the usability of the result, while too little cannot provide sufficient security. Sensitivity is a key parameter that determines the amount of noise added. It refers to the largest change to the query result caused by adding or deleting any record in the data set.

For $f: D \rightarrow R^d$, the L1-sensitivity of $f$ is

$$\Delta f = \max_{D_1, D_2} ||f(D_1) - f(D_2)||_1 \quad (4)$$

for all $D_1$, $D_2$ differing in one element at most.

The sensitivity of a function is determined by the function itself, and different functions will have different sensitivities. For functions with lower sensitivity, sufficient privacy protection can be achieved with the addition of only a small amount of noise. However, for some sensitive functions (such as the median function), it is required to add a lot of noise to achieve the same level of protection.

The most common implementation mechanisms are Laplace Mechanism and Exponential Mechanism. Probability Density Function (PDF) for a random variable with Laplace distribution is defined as follow:

$$Laplace(x|\,\mu, b) = \frac{1}{2b}\exp(-\frac{|x-\mu|}{b}) \quad (5)$$

Let $b = \frac{\Delta f}{\epsilon}$ where f is the query function. Then, we have

$$Laplace(x|\,\mu, \epsilon, \Delta f) = \frac{\epsilon}{2\Delta f}\exp(-\epsilon\frac{|x-\mu|}{\Delta f}) \quad (6)$$

## IV. Results Analysis

### A. *Efficiency and accuracy evaluation*

The approach used for evaluation has two considerations body temperature (BT) and heart rate (HR). It measured time intervals (minutes) of sensing frequency to analyze the differences of inferencing results. Fine and coarse inference algorithms are applied to show the differences and efficiency of each of these cases. Dataset [12] was used for HR and BT with Matlab R2019b for the inference algorithm.

Evaluation results have been displayed in Fig. 2 and depict body temperature (BT) and heart rate (HR) sensed on a minute basis. In applying a 1% and 2.5% inference rate to BT and HR data respectively, the volume of data to be transferred was reduced by 76% for BT and 73% for HR.

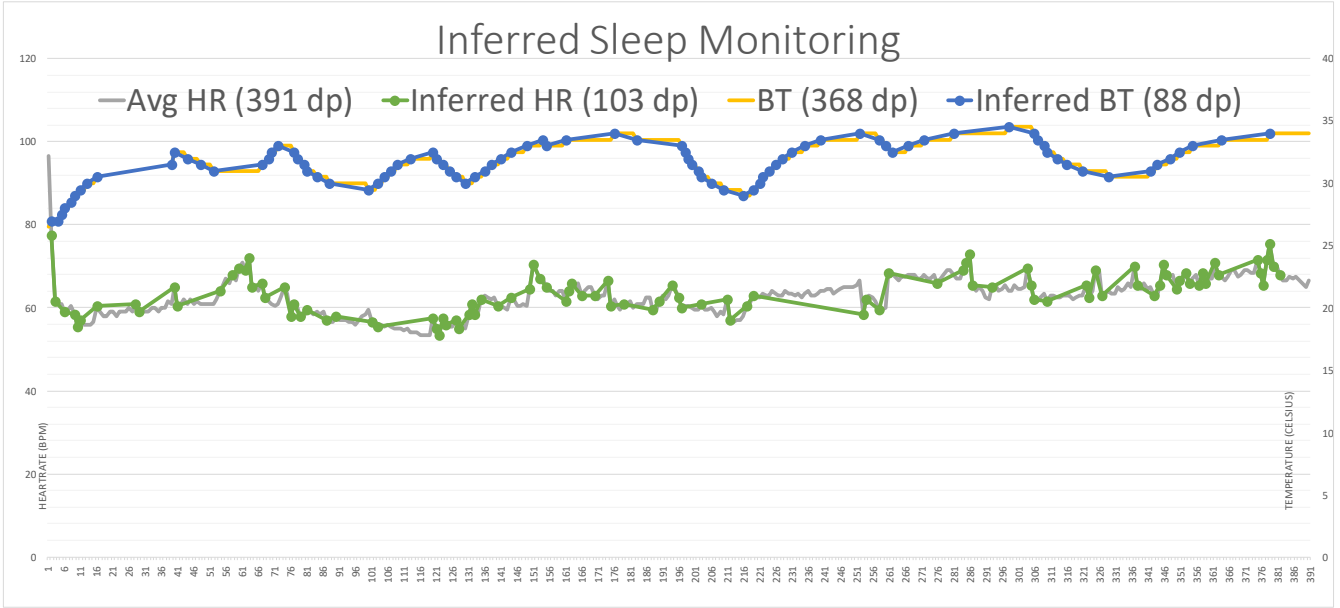

Fig. 2. Inferred HR and BT of sleep monitoring data (minutes) – inferred BT data represents the original well whilst inferred HR data shows relatively more gaps (this could be improved by using beacon data sampling)

BT inference shows better results representing almost identical data as opposed to HR. In other words, whilst the data savings rates are similar for both BT and HR, the accuracy of inference in both types of data were very different and could be reflective of the inherent differences in what these data are measuring. Distortion of the original data can occur when an inference system disregards data to transmit if it does not vary sufficiently from previous or adjacent DPs i.e. does not meet a stated VR threshold and is therefore determined to be of little significance and considered unnecessary to be transmitted. This distortion is especially so for data that are measured at shorter intervals, as the data could be trending over a longer term but simply due to the shorter frequency of data measurements, do not have time to vary significantly between each subsequent data measurement. This limitation has also been discussed in greater detail earlier in Section III along with a potential solution, which is to add DPs which function as beacons. These beacons transmit data at set intervals regardless of whether they meet the VR threshold criteria, and therefore helps to maintain the accuracy of the overall inference data without compromising heavily on data savings. In these experiments, beacon DPs were set to minute intervals. A finer inference VR threshold can provide greater accuracy; however, it results in lesser transmission savings and decreases the overall efficiency rate from the perspective of data transmission. Certain situations may simply require a general idea of the trend rather than valuing exact or accurate figures – in these cases, a coarser inference VR method could be used instead which places greater priority on data saving. The exact interval of beacon DPs would depend upon the context and solution or application requirements for which this inference is being implemented.

### B. *Privacy-preserving evaluation*

This part of the paper aims to study the efficiency of the proposed scheme from privacy preservation perspective. Research on privacy-preserving approaches in eHealth clouds have commonly tended to focus on cryptographic methods such as symmetric key encryption (SKE) and attribute-based encryption (ABE) [13]. In order to prove the efficiency of the proposed data inference framework for IoHT, the correlation between plaintext size and crypto texts size can be tested in two main categories: SKE and ABE. For SKE evaluation, the simulation is conducted in OnlineDomainTools [16] for three main symmetric encryption techniques: advanced encryption standard (AES), data encryption standard (DES), and blowfish.

TABLE I: Data Savings for Different Variance Rates (24 hour samples)

| VR | 0% | 2.5% | 5% | 10% | 20% |
|---|---|---|---|---|---|
| DP | 1420 | 691 | 306 | 146 | 17 |
| Saving (%) | N/A | 51.3 | 78.5 | 89.7 | 98.8 |
| Accuracy result | N/A | Very Good | Useable | Poor | Unusable |

Over a course of 24 hours in an experiment, a total of 1420 heart rate DPs were sensed and processed at various inference rates ranging from 2.5%, 5%, 10% and 20% VRs. After inferencing algorithms are applied to the data, the number of DPs to be transferred were reduced significantly as shown in TABLE I. VRs ranged from 0%, 2.5%, 5%, 10% and 20% which resulted in savings ranging from 0%, 51.3%, 78.5%, 89.7% and 98.8% respectively. Plaintext size for 0% savings can be considered as 1024B (1MB). Plaintext size for other degree of savings can be obtained by:

$$plaintext\ size\ (B) = \frac{(100 - saving) * 1024}{100} \quad (7)$$

The impact of varying the plain text size is shown in Fig. 3. The mode is set to ECB (electronic code book) while maintaining the key at 128 and evaluating varying plain text sizes from 1024, 498, 220, 105, and 12 with AES, DES, and blowfish encryption functions. The results show that as the size of the plain text size decreases, the size of crypto text also decreases accordingly. Comparing TABLE I and Fig. 3, when VR is equal to 2.5% and the accuracy result is very good, the crypto text size for 1024 bytes of data is equal to 496 bytes. However, there is no clear evidence on whether one encryption technique was better than another.

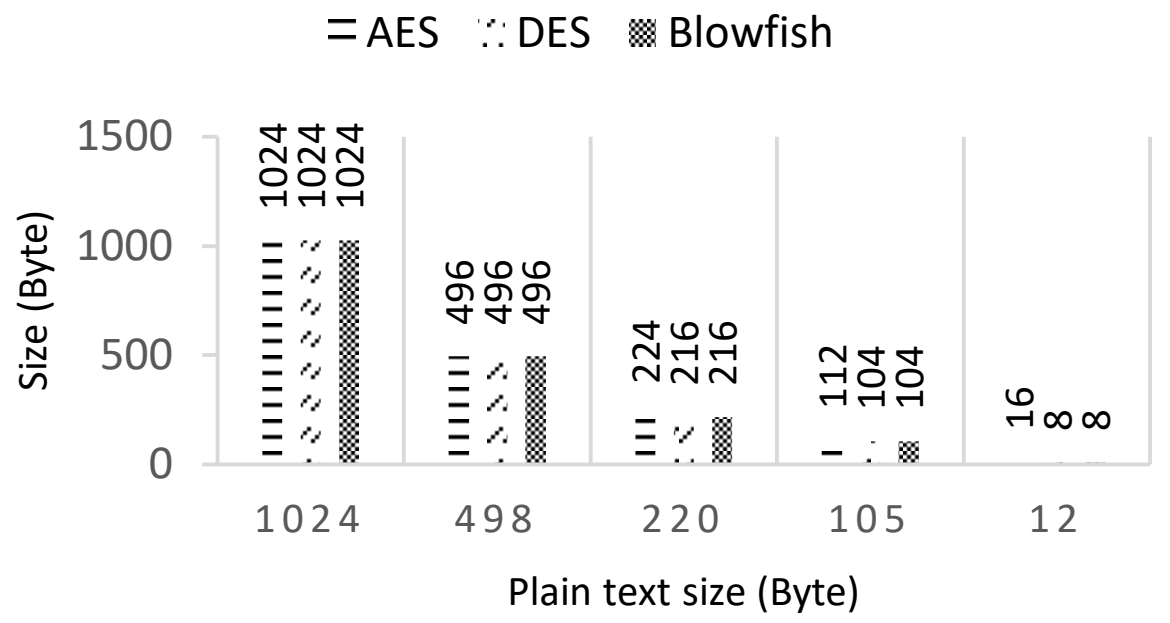

Fig. 3: Evaluation of varying plain text size and its effect on crypto text size

We have also evaluated the effectiveness of our solution by utilizing differential privacy. The dataset used in the experiment contains information about body temperature, gender, and heart rate for 130 people.

In the proposed model in this paper, data will eventually be used for statistical queries. For example, the average heart rate of someone in a day will be queried. The difference between one more record and one less record on the statistical results is defined as the sensitivity of the query algorithm, denoted as Δf. In order to provide $\epsilon$-differential privacy protection for our data, the output result will be:

$$Out_Result = Real_Result + \text{Laplace}(\frac{\Delta f}{\epsilon}) \quad (8)$$

The Laplace (Δf/ $\epsilon$) is the Laplace noise which was added to protect the real data. According to Eq. (8), the sensitivity of our query algorithm is first required to be analyzed, from which, an appropriate $\epsilon$ to obtain the required Laplace noise is selected. By definition of sensitivity, it is logical to infer that the greater the sensitivity, the greater the noise, and the smaller the sensitivity, the smaller the noise. An assumption is made that Δf = 1 (that is, the addition of each new record will cause the result to change by 1, which is very large). Therefore, the following experiments are conducted with a sensitivity of 1 (Δf = 1), and the distribution of Laplace noise added to the data is equal to Laplace (1/ $\epsilon$). Noise is added to satisfy the Laplace (1/ $\epsilon$) distribution to each heart rate data in the original data set. Six experiments were performed where $\epsilon$ was set equal to 0.01, 0.05, 0.1, 0.2, 0.5, 1.0 and the results compared to observe how differential privacy protects the original data. Following this, the average values of the original data were identified and compared with the original data statistics to compare the performance of differential privacy.

Fig. 4 shows the experimental results under six $\epsilon$. The x-axis of each sub-figure in Fig. 4 represents the index of the DP in the data set. The y-axis represents the heart rate value of this DP. The blue line in the figure represents the heart rate value in the original data set, and the red point represents the value after the addition of Laplace noise to each DP in the original data. The distribution of Laplace noise added are Lap (1/0.01), Lap (1/0.05), Lap (1/0.1), Lap (1/0.2), Lap (1/0.5), and Lap (1/1).

TABLE II: QUERY RESULTS UNDER DIFFERENT $\epsilon$

| The value of $\epsilon$ | 0.01 | 0.05 | 0.1 | 0.2 | **0.5** | 1.0 |
|---|---|---|---|---|---|---|
| Average value of raw data | 73.76 | 73.76 | 73.76 | 73.76 | **73.76** | 73.76 |
| Average value of privacy-preserved data | 73.84 | 73.18 | 73.14 | 73.89 | **73.76** | 73.76 |

Based on the trend in changes of the sub-figures, it can be observed that with the increase of $\epsilon$, the added noise begins to decrease i.e. the degree of deviation of red points from the blue line begins to decrease. When $\epsilon$ = 1, the noised data almost coincides with the original data. According to the trend of Fig. 4, it can be observed that when $\epsilon$ is smaller (as in the sub-figure with $\epsilon$ = 0.01), the degree of privacy protection provided by random algorithms is greater. Conversely, when $\epsilon$ is larger (as in the sub-figure with $\epsilon$ = 1), the degree of privacy protection provided by random algorithms is lower.

Each DP after adding Laplace noise will deviate from the original data to a certain extent. However, this is not necessarily important as users in practice may not query a specific value such as their heart rate at a specific point, but may be more concerned about the average value over a certain period of time. The average value of both the original data set and the noised data set in all six experiments were calculated and the results summarized in TABLE II. It can be seen from TABLE II that the size of the noise added to the original data set is different in each experiment. The statistical results (the statistical results after adding noise) deviate from the true statistical results (the raw data statistical results) to different degrees. The smaller the deviation, the higher the availability of data. When $\epsilon$ = 0.01, $\epsilon$ = 0.05, $\epsilon$ = 0.1, $\epsilon$ = 0.2, there is a relatively large deviation level, and the availability of data is low. When $\epsilon$ = 0.5, the degree of deviation is very small (almost close to 0) and the data availability is high. The purpose of adding Laplace Noise is to ensure the availability of data while protecting user privacy. The experimental results from Fig. 4 and TABLE II show that there is a compromise in privacy protection and data availability – obtaining greater results in one requires compromising the other. In comparing Fig. 4 and TABLE II under these considerations, data protection capability and data availability were the best when $\epsilon$ = 0.5.

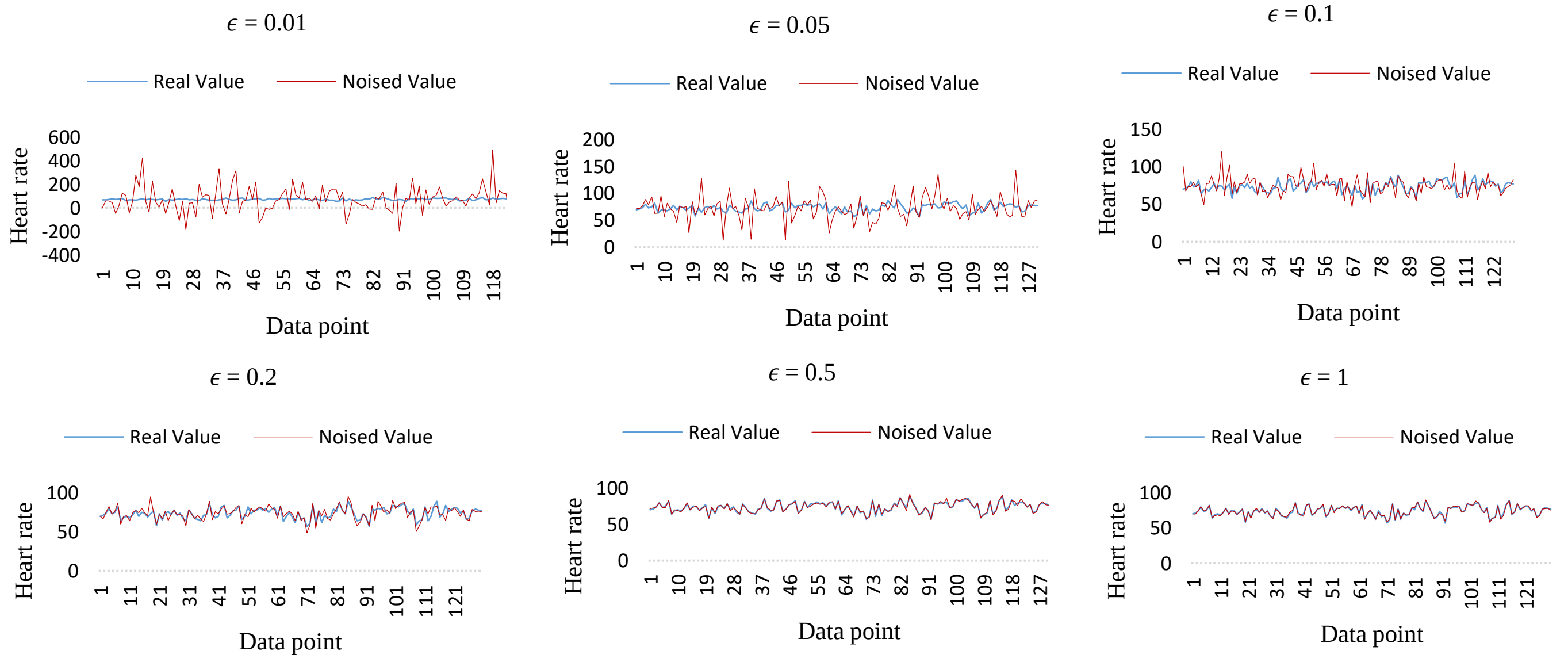

Fig. 4. Evaluation of differential privacy with six different values of $\epsilon$

## V. Conclusion

Energy efficiency and privacy preservation of sensitive health data are essential in IoHT networks, which consist largely of smart devices limited by battery constraints. In this paper, a two-tier data inference framework has been proposed to conserve energy consumption by reducing unnecessary data transmission within the IoHT network while still maintaining high accuracy. The results suggest that applying 1% to 2.5% variance rate by the inference system achieved the best accuracy. It was also shown that this amount of VR decreases nearly half of the crypto text size using main symmetric encryption techniques. Another major finding was that applying differential privacy with a $\epsilon = 0.5$ satisfies data protection and data availability requirements. The experimental results show that the proposed system is beneficial for saving energy of IoT devices and security analysis suggests that the differential privacy technique can protect against sensitive health data from being obtained maliciously. In our future work, we will look into how to incorporate highly efficient blockchain and federated learning techniques [17,18] into our solution to improve privacy preservation while maintaining high accuracy in data inference.